# Geometric origin of macroscopic alignment in granular flows

Christopher Harper [*]
*Department of Earth Sciences, University of Oregon, Eugene, Oregon 97401, USA*

Eric C. P. Breard and PJ Zrelak
*School of GeoSciences, University of Edinburgh, Edinburgh, United Kingdom and Department of Earth Sciences, University of Oregon, Eugene, Oregon 97401, USA*

George W. Bergantz
*Department of Earth and Space Sciences, University of Washington, Seattle, Washington 98195, USA*

(Received 1 June 2026; accepted 24 August 2026; published 17 September 2026)

Predicting the nematic alignment of nonspherical particles in sheared granular flows is essential for understanding the rheology, packing, and constitutive response of dense particulate media. While macroscopic fabric is typically attributed to complex multibody interactions, stress transmission, and dissipative collisions, empirical observations reveal that the steady-state nematic order parameter $S_2$ depends primarily on particle aspect ratio and remains remarkably insensitive to shear rate and interparticle friction. Here, we show that this leading-order alignment emerges directly from single-particle boundary geometry without resolving dynamical equations of motion. Assuming uniform contact probability along a particle perimeter, we derive an analytical transform linking local boundary curvature $\kappa(\theta)$ to the distribution of contact normals, $P(\theta) \propto 1/\kappa(\theta)$, which in turn geometrically constrains the phase space of admissible particle orientations. This minimal framework accurately predicts the magnitude of $S_2$ across the full continuum of aspect ratios for smooth ellipsoids as well as the singular limit of faceted rectangles and cylinders. Our analytical predictions capture the envelope of three-dimensional discrete element simulations and match laboratory measurements on sheared rice grains and glass cylinders across decadal variations in shear rate. By identifying particle geometry as the primary control parameter for granular alignment, this work provides a first-principles physical foundation for geometric saturation at the critical state, establishing a universal baseline upon which dynamical and frictional effects act as secondary modulations.

## I. INTRODUCTION

The alignment of non-Brownian, nonspherical particles in flow is widely observed in both natural [1–5] and industrial settings [6–8]. This nematic alignment is often referred to as a material "fabric," or shape-preferred orientation (SPO). This anisotropic alignment arises in both suspensions and (dry) granular flows, and has been the focus of study from dilute to particle-rich (dense) concentrations [9,10]. The emergence of fabric can change the packing, strength, and permeability of a multiphase mixture, as it influences Reynolds dilation, coordination number, anisotropic force distributions, force-chain orientation, and bulk rheology [11–19]. Hence, predicting the terminal orientational and positional ordering in a multiphase mixture of nonspherical particles has significant implications for industrial design and the interpretation of geophysical processes.

It is well established that a particle-rich mixture (dense) under shear will eventually self-organize to what has been called a critical state, where continued shear strain yields no further changes in the volume, shear strength, coordination number, or shape-preferred orientation of the system [20]. This has been demonstrated both experimentally, using material as diverse as rice, grains, glass cylinders, and fibers [21–23], and in numerical simulations [9,24–26]. One property of the critical state is the emergence of a persistent uniaxial nematic ordering. The strength of the ordering is expressed by the order parameter $S_2$, which is the largest eigenvalue of the symmetric traceless order tensor $Q$ [Eq. (7)] [9,21].

A value of $S_2$ equal to unity means all the particle are aligned, and a value of zero means the alignment is isotropic with no preferred orientation. Empirical evidence shows that $S_2$ is strongly dictated by particle aspect ratio and is remarkably insensitive to realistic values of particle friction [25] or shear rate [21]. While it has been elegantly proposed that this critical state represents a geometrically "saturated" fabric constrained by steric exclusions within the contact network [27], the precise value of $S_2$ has remained an empirical observation. Despite its reproducibility across various families of shapes, from ellipsoids to flat disks and elongated rods [25], there is currently no first-principles framework to predict the terminal alignment for a given particle assemblage [9]. This work offers such a framework by examining the relationship between two alternative constructions of granular fabric.

[*]Contact author: charper4@uoregon.edu

Thus far, we have used the term "fabric" to refer to the distribution of particle orientations. More generally, however, granular fabric may also be characterized by the distribution of contact normals within the contact network. These descriptions are not interchangeable. A given contact normal does not uniquely determine the orientations of the particles forming the contact, as an infinite set of mutual particle orientations can generate the same local geometry. Consequently, particle orientations and contact-normal fabric encode different aspects of the granular microstructure.

Recent work by Wang *et al.* [28] highlighted the importance of this distinction by examining the evolution of both normal-contact- and particle-orientation-based fabric tensors. They showed that the contact-normal fabric evolves more rapidly and is more directly coupled to dilatancy. While this elucidates the role of contact-normal-derived fabric as the primary fabric state variable in their anisotropic critical state theory (ACST), it does not dismiss the importance of particle orientations. In fact, their findings emphasize the role particle orientation plays in affecting the evolution of the contact network, which ultimately controls the constitutive response. This manifests in Wang *et al.*'s final description of granular fabric including a negative exponential that effectively slows the evolution of the contact-normal-derived fabric tensor based on the state of the particle orientation fabric. This is an empirical link; here, we offer a direct probabilistic link constructed from particle geometry.

Although multiple particle orientations can produce the same contact normal, these configurations are not equally probable. Regions of high boundary curvature contribute disproportionately to the distribution of contact normals, whereas regions of low curvature contribute comparatively little. Our central hypothesis is that the steady-state contact-normal fabric established at the critical state therefore imposes a geometric constraint on the distribution of particle orientations. This constraint becomes increasingly restrictive for particles with more strongly localized curvature (e.g., ellipses of increasing aspect ratio), naturally producing the observed increase in the orientational order parameter $S_2$ with aspect ratio. Importantly, this relationship emerges from particle geometry alone and is independent of the complex force-chain dynamics or dissipative interactions typically associated with granular flow. The resulting framework provides a physically grounded geometric baseline for understanding the statistical emergence of granular fabric, reconciling the structural constraints imposed by the contact network with the fundamental geometry of the constituent particles.

## II. THEORETICAL FRAMEWORK: MAPPING BETWEEN PARTICLE GEOMETRY AND LIKELIHOOD OF PARTICLE NORMALS

In this section, we establish a purely geometric derivation for the distribution of contact normals in a granular assembly. By assuming that the likelihood of potential particle-particle contacts is uniformly distributed along a particle's perimeter, we derive a minimal description for how boundary curvature influences the resulting distribution of normal vectors along the boundary. This approach isolates the role of particle geometry from the dynamical complexities of force transmission and dissipation, allowing us to show that this purely geometric constraint captures the dominant features of the observed orientational statistics.

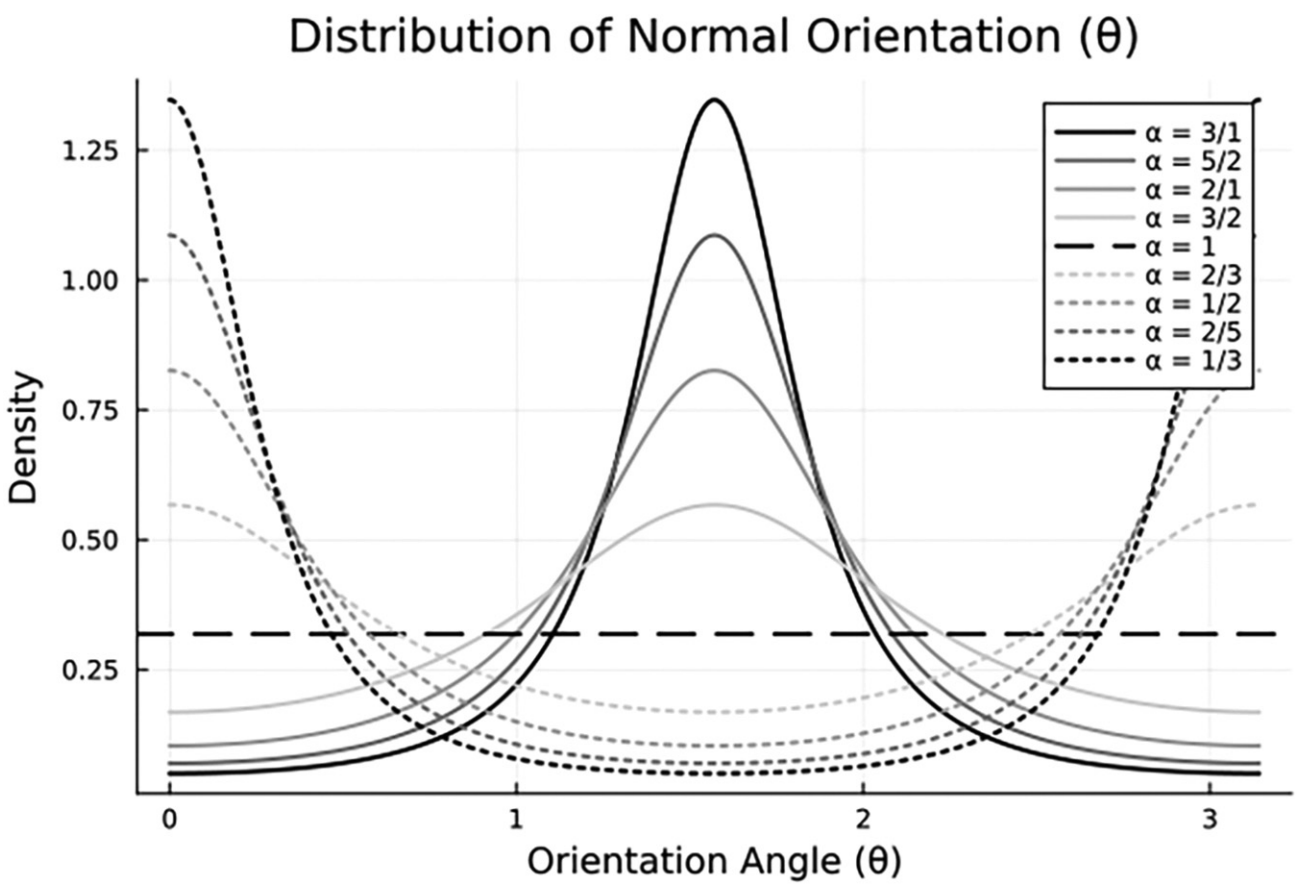

FIG. 1. Geometric warping of contact-normal orientations by particle shape. For a circular particle ($\alpha = 1$), uniform sampling along the boundary yields a uniform distribution of contact-normal orientations $P(\theta)$. As the aspect ratio deviates from unity, the distribution becomes more biased.

To comprehensively account for the role of geometry, we make three simplifying assumptions:

(1) *Particle shape*. Particles are convex, with boundaries approximated by either smooth (ellipsoids) or piecewise-smooth (rectangles) geometries.

(2) *Planar confinement*. Under steady simple shear, alignment occurs predominantly within the shear plane; out-of-plane fluctuations are neglected.

(3) *Geometric control*. Within the shear plane, orientation statistics are determined primarily by geometry, with dynamics entering only through contact formation.

Given the assumptions above, the role of particle boundary curvature in setting the contact distribution simplifies to the two-dimensional perimeter curvature within the shear plane. Contact locations are then modeled as uniformly distributed with respect to this perimeter's arc length $s$, such that the probability that any point is a contact point, $p(s)$, is constant over the perimeter. Importantly, this uniform distribution of particle contacts does not imply a uniform distribution of normal orientations (see Fig. 1). It is precisely this fact that points sampled uniformly from a boundary can exhibit preferred orientations, which lies at the heart of our argument.

Each point along the boundary is uniquely and invertibly associated with an outward normal, whose direction we parametrize by the polar angle $\theta$ in the shear plane. This uniqueness, and invertibility, is guaranteed by the shape's strictly convex nature: The normal angle $\theta$ must increase monotonically as one traverses the perimeter. The map from arc length $s$ to orientation angle $\theta$ defines a change of variables between probability measures. Applying the standard relation,

$$P(\theta) = p(s)\left|\frac{ds}{d\theta}\right|,$$

we see that the orientational distribution is governed by the geometric factor $|ds/d\theta|$, which measures how much boundary length contributes to a given angular interval of the normal.

This quantity is directly related to the curvature $\kappa$ of the boundary,

$$\frac{ds}{d\theta} = \frac{1}{\kappa}.$$

Thus,

$$P(\theta) \propto \frac{1}{\kappa(\theta)}. \tag{1}$$

To evaluate this explicitly, we consider an ellipse with semiaxes $a$ and $b$; the points along the boundary are parametrized by

$$r(\phi) = (a\cos\phi,\ b\sin\phi),$$

where $\phi$ is the standard parametric angle. The curvature as a function of $\phi$ is

$$\kappa(\phi) = \frac{ab}{(a^2\sin^2\phi + b^2\cos^2\phi)^{3/2}}. \tag{2}$$

Each point also defines a unique outward normal. As before, $\theta$ denotes the polar angle of this normal. The geometric relation between $\phi$ and $\theta$ is

$$\theta = \arctan\left(\frac{a}{b}\tan\phi\right).$$

Using this transformation, we express the curvature with respect to the normal orientation:

$$\kappa(\theta) = \frac{(a^2\cos^2\theta + b^2\sin^2\theta)^{3/2}}{a^2b^2}. \tag{3}$$

In this form, we have captured the desired role of curvature as a geometric weight on the likelihood of various contact normals.

Figure 1 illustrates how $P(\theta)$ varies with aspect ratio. With the analytical form $P(\theta) \propto 1/\kappa(\theta)$, we generate predicted orientation distributions for ensembles of ellipsoidal grains, including both prolate and oblate shapes.

To demonstrate the generality of the curvature-based framework, we consider the limiting case of a rectangular particle in the shear plane. In contrast to smooth shapes such as ellipses, a rectangle consists of flat edges separated by sharp corners. Along each flat edge, the outward normal direction is constant, while at the corners the curvature is infinite.

Because contact normals are defined up to a sign, we restrict orientations to $\theta \in [0, \pi)$, identifying $\theta$ and $\theta + \pi$. Under this convention, a rectangle with edge lengths $w$ (horizontal) and $h$ (vertical) admits only two distinct normal orientations:

$$\theta \in \left\{0,\ \frac{\pi}{2}\right\}.$$

Assuming uniform sampling of contact locations along the boundary, the probability of observing a given orientation is proportional to the total length of edges sharing that normal. With total perimeter $L = 2w + 2h$, this yields the discrete distribution

$$P(\theta) = \begin{cases} \frac{2h}{L}, & \theta = 0, \\ \frac{2w}{L}, & \theta = \frac{\pi}{2}, \\ 0, & \text{otherwise.} \end{cases} \tag{4}$$

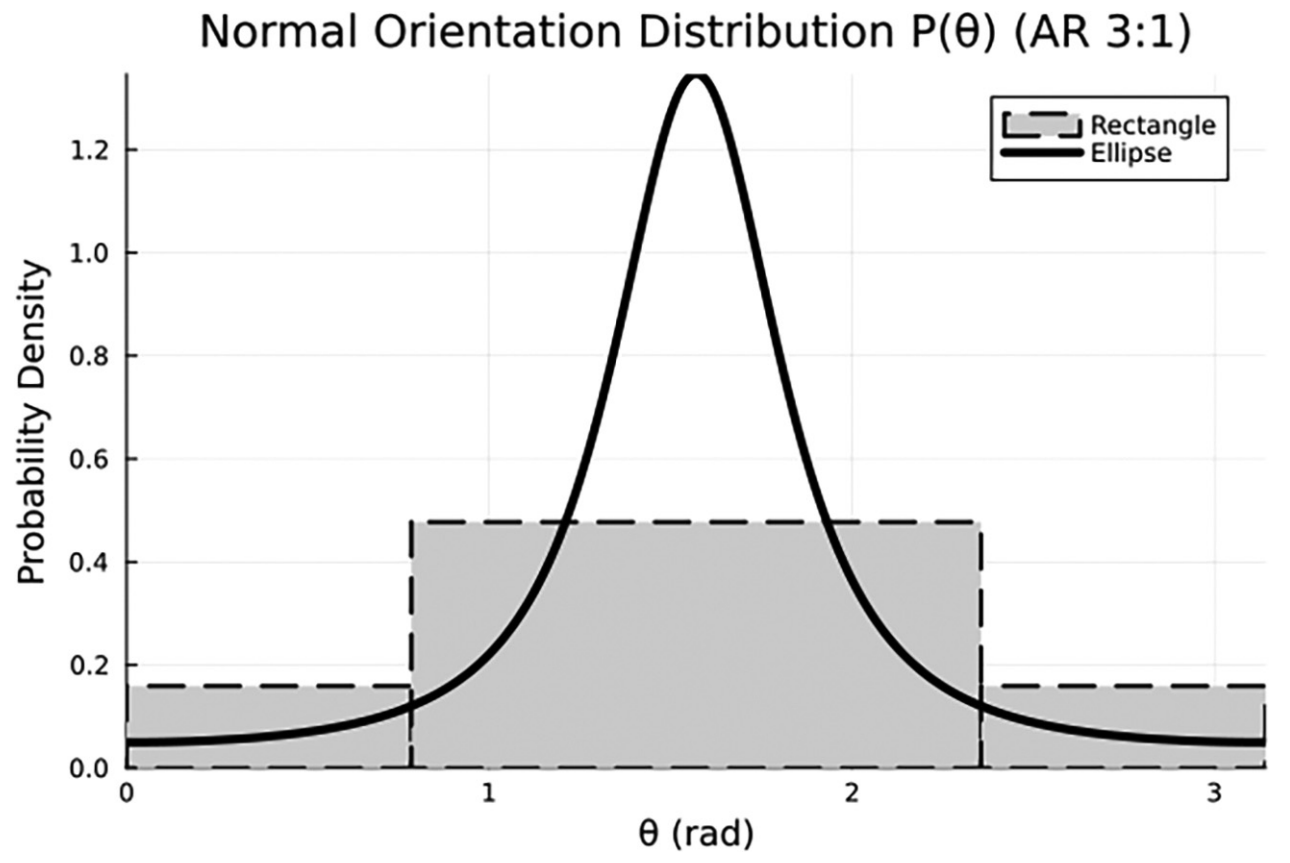

FIG. 2. Comparison of normal orientation distributions for smooth vs piecewise-flat boundaries. The solid curve represents the continuous probability density $P(\theta)$ for an ellipsoid with an aspect ratio of 3:1. The shaded regions represent the distribution for a rectangle of the same aspect ratio.

To better understand in what ways rectangles represent a limiting form of the ellipsoid geometry, we represent the distribution as a sum of Dirac delta functions:

$$P(\theta) = \frac{2h}{L}\,\delta(\theta) + \frac{2w}{L}\,\delta\left(\theta - \frac{\pi}{2}\right), \qquad \theta \in [0, \pi). \tag{5}$$

This form makes the orientation distribution's concentration at the face normals explicit, weighting it by the relative edge lengths. It can be understood as the singular limit of the continuous relation $P(\theta) \propto 1/\kappa(\theta)$: along the flat edges $\kappa \to 0$, leading to a concentration of probability, while at the corners $\kappa \to \infty$, yielding negligible contribution. To illustrate this geometric collapse, Fig. 2 contrasts the continuous orientation distribution of an ellipsoid with the discrete probability mass of a rectangle of the same aspect ratio (3:1). For the rectangle, the shaded domains do not represent a continuous spread of normal angles; rather, the width and area of the shaded bins represent the total probability mass (proportional to the flat edge lengths) that collapses entirely onto the discrete face normals (e.g., the long edge mapping exactly to $\theta = \pi/2$).

The rectangular case emerges as a limiting form in which a continuous spectrum of orientations collapses onto a discrete set. Although a pathological extreme of our curvature argument, such piecewise-continuous geometries are physically relevant, appearing in the rectangular cross sections of cylindrical simulations and the faceted surfaces of fragmented natural materials.

While these distributions are derived from a single-particle geometric model, our hypothesis requires the extension of these distributions to dense systems. The following section details this extension.

## III. METHODS: CONSTRUCTING THE STATISTICAL ENSEMBLE FOR MACROSCOPIC ALIGNMENT

For any ellipsoidal particle, we generate the analytical expression for $\kappa(\theta)$ [Eq. (3)] from the major and minor axes of the ellipse formed by projecting the ellipsoid onto the shear plane. The probability density function $P(\theta)$ is the normalized reciprocal of this function, as shown in Eq. (6):

$$P(\theta) = \frac{f(\theta)}{\int_0^{\pi} f(\theta)\,d\theta}, \quad \text{where} \quad f(\theta) = \frac{1}{\kappa(\theta)}. \tag{6}$$

While $P(\theta)$ is not a standard distribution, its exact analytical form makes it an ideal candidate for rejection sampling, whereby random candidate angles are uniformly generated and accepted or discarded based on whether an independent uniform random variable falls below $P(\theta)$. We generate 2000 orientations sampled from $P(\theta)$ through rejection sampling, denoted as the set $\Theta$. The sample size of 2000 is somewhat arbitrary, and reflects sampling one contact normal per particle in the Bilotto simulations to which we compare our results. Sampling for rectangular particles is similar; however, we directly simulate the uniform perimeter-based sampling arising from Eq. (4).

To quantify macroscopic alignment, we construct the orientation tensor following standard conventions for granular nematics [9,21]. For each sampled angle $\theta_n \in \Theta$, we define the two-dimensional unit normal vector $\hat{u}_n = (\cos\theta_n, \sin\theta_n)$ in the shear plane. The macroscopic orientation tensor $Q$ is computed as the ensemble average over all $N$ sampled contacts:

$$Q = \frac{1}{N}\sum_{n=1}^{N}\left(\hat{u}_n \otimes \hat{u}_n - \frac{1}{2}I\right), \tag{7}$$

where $I$ is the $2 \times 2$ identity matrix and $\otimes$ denotes the outer product.

Because $Q$ is a real, symmetric tensor, we determine its principal directions of alignment by computing its eigenvalues. The eigenvector associated with the largest eigenvalue represents the dominant macroscopic orientation. Let $\lambda_{\max}$ be the maximum eigenvalue of the ensemble-averaged $Q$ tensor; scaling $\lambda_{\max}$ by 2 yields the uniaxial nematic order parameter $S_2$:

$$S_2 = 2\lambda_{\max}\,. \tag{8}$$

The numerical sampling described above effectively fixes a contact normal along a constant direction $\theta_{\text{lab}}$ in the laboratory frame, randomly assigning boundary points to this contact orientation. As established in Sec. II, the bijection between boundary position and outward normal allows each sampled point to be associated with an intrinsic normal orientation $\theta \sim P(\theta)$ measured relative to the particles primary axis. The use of intrinsic here emphasizes that $\theta$ depends only on the particle shape and is independent of the laboratory's reference frame. In the laboratory frame, the orientation of the contact normal $\theta_{\text{lab}}$ is the composite sum of the particle's primary-axis orientation in the laboratory $\psi_{\text{lab}}$ and its intrinsic contact-normal angle $\theta$:

$$\theta_{\text{lab}} = \psi_{\text{lab}} + \theta \rightarrow \psi_{\text{lab}} = \theta_{\text{lab}} - \theta. \tag{9}$$

This relation is crucial because laboratory experiments and numerical simulations typically measure particle orientation according to $\psi_{\text{lab}}$. Because $\theta_{\text{lab}}$ is arbitrary but fixed, Eq. (9) denotes a rigid coordinate rotation between the intrinsic contact normal being sampled and the laboratory orientation of the primary axis. While a rigid rotation changes the orientation of the $Q$ tensor's eigenvectors relative to the laboratory frame, it leaves its eigenvalues (principal values) unchanged. Therefore, the order parameter $S_2 = 2\lambda_{\max}$ calculated from the distribution of contact normals, $P(\theta)$, is mathematically identical to the $S_2$ of the associated particle's primary axis as measured in the laboratory frame.

Furthermore, we can reconstruct the intrinsic director of the particle ensemble from the fact that the mode of $P(\theta)$ corresponds to the broadest, lowest-curvature flanks of the perimeter ($\theta = \pi/2$), which aligns the dominant contact normal perpendicular to the particle's primary axis. Because our minimal geometric framework samples single-particle geometry without imposing external velocity gradients or stress fields, it predicts the magnitude of orientational alignment ($S_2$) around this director rather than prescribing an absolute tilt angle in the laboratory frame (where alignment is modulated by gravity, shear kinematics, or principal stress axes). Fixing a shared contact angle in our simulations allows for the measurement of the order parameter as an intrinsic geometric property of the particle shape. Our results in the next section will demonstrate that this minimal baseline captures the leading-order behavior of alignment across diverse critical-state dense granular systems. Our thoughts as to why the particle geometry plays such a leading role in controlling alignment when so many other physical mechanisms are present are saved for the discussion.

## IV. RESULTS

We apply our geometric framework to predict the uniaxial nematic order parameter $S_2$ across a wide range of ellipsoidal aspect ratios ($\alpha \in [0.1, 10]$). These predictions are compared against three-dimensional, frictional discrete element method (DEM) simulations from Bilotto *et al.* [9] and laboratory experiments using rice grains by Börzsönyi *et al.* [21]. To facilitate comparisons across different studies, we convert all aspect ratios ($a/b$) into a shape ratio ($R_g = (a-b)/(a+b)$).

As shown in Fig. 3, the geometric model closely follows the results of highly frictional systems ($\mu = 1.0$) and successfully captures the general envelope of the DEM data of Bilotto *et al.* [9]. The experimental alignment of rice grains also aligns well with our geometric baseline, despite decadal variations in shear rate. These results suggest that while friction and kinematics modulate the magnitude of alignment, the first-order behavior of granular fabric is fundamentally dictated by the warping of contact probability due to boundary curvature.

When considering a different class of particle geometries (disks and rods with planar surfaces), the results are less clear. For these particles, we utilize the rectangular model derived in Eq. (5). As illustrated in Fig. 4, the rectangular cross section better describes disks and rods of intermediate to low aspect ratios; however, for more extreme aspect ratios, the simulations almost snap onto the elliptical predictions.

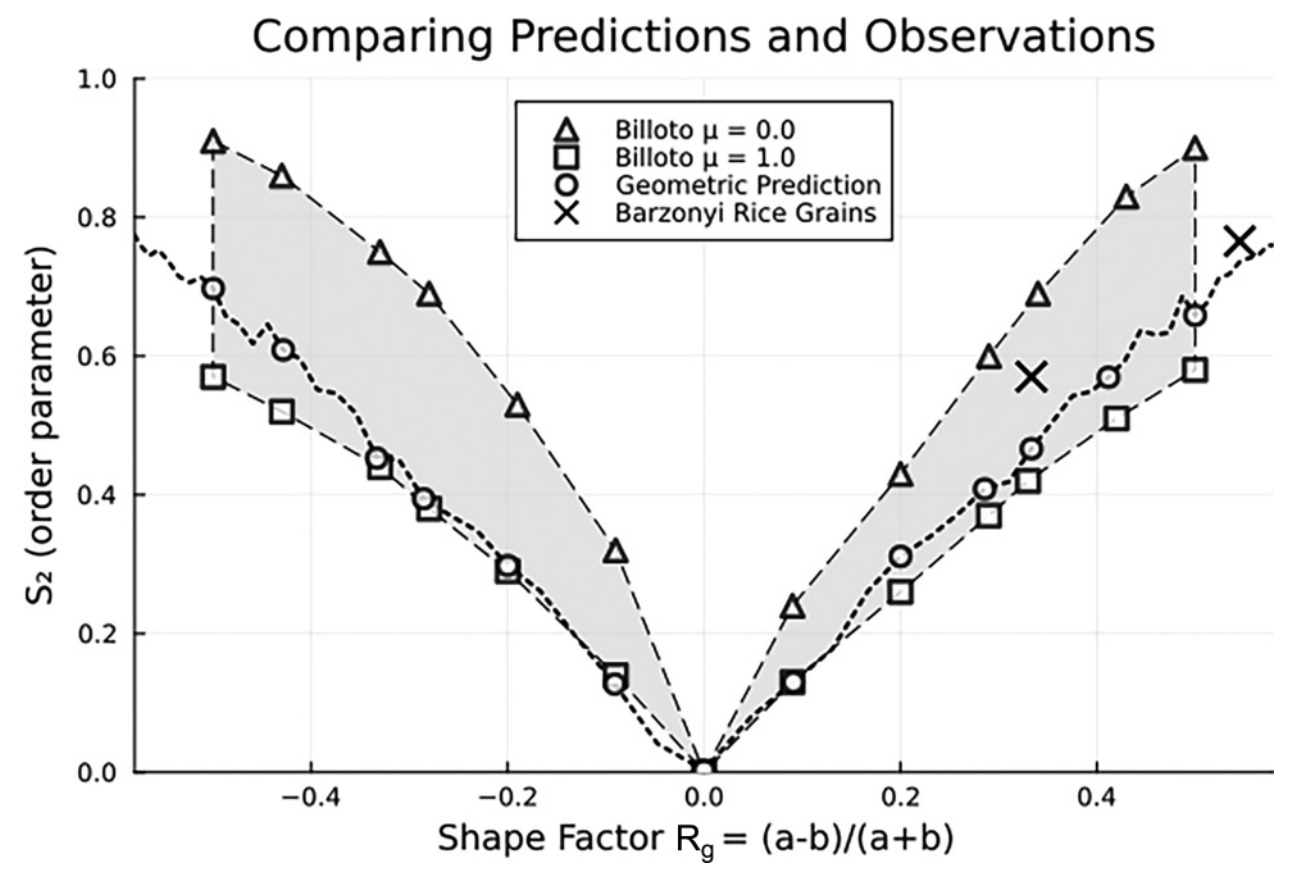

FIG. 3. Comparison of geometric predictions with numerical and experimental data. The uniaxial nematic order parameter $S_2$ is plotted against the shape ratio $R_g = (a-b)/(a+b)$. Our analytical prediction (dotted line) identifies specific values for ellipsoidal geometries (circles). The shaded region represents the envelope of outcomes from DEM simulations by Bilotto *et al.* [9], ranging from frictionless ($\mu = 0.0$, upper bound) to highly frictional ($\mu = 1.0$, lower bound) systems. Crosses (X) denote experimental results for rice grains of aspect ratios 2.0 and 3.4, averaged across shear rates from 0.01 to 1.0 $s^{-1}$ (adapted from Ref. [21]).

## V. DISCUSSION

The emergence of a stable shape-dependent alignment in the critical state has previously been described as a "geometric saturation." In this framework, Radjaï *et al.* [27] proposed that the contact anisotropy is bounded by available spaces of accessible fabric states; however, the precise value of this saturation point has remained an empirical observation dependent on specific material properties. Our framework provides a physical basis for these states by demonstrating that the steady-state distribution of contact normals is mapped directly from the particle's boundary curvature. By identifying the geometric weight $1/\kappa(\theta)$ as the primary driver of contact normal density, we provide a predictive link between particle geometry and these fabric limits, effectively grounding "accessible states" in the fundamental curvature of the particle boundary.

This perspective also clarifies how the order parameter as measured from the particles' intrinsic shape captures the leading behavior of an order parameter influenced by processes playing out in a global laboratory frame. In a sheared dense granular system, the laboratory orientation of any contact normal is the composite sum of the particle's orientation in the flow and its local boundary normal angle. Because continuous shear drives the dense contact network to geometric saturation, particle rotation is physically bounded by the geometry of available contact tangents. Under these critical-state conditions, the macroscopic particle-orientation order parameter $S_2$ saturates directly to the single-particle geometric limit, explaining why our minimal framework captures the terminal alignment observed across diverse numerical and experimental systems.

Recent work by Marschall *et al.* [29] examined the surface probability distribution of contacts $P(\vartheta)$, in sheared assemblies of frictionless ellipsoids and spherocylinders, demonstrating that contacts preferentially localize along the narrowest particle widths near jamming. Their results provide an important complementary perspective to the present framework. In their formulation, $P(\vartheta)$ characterizes the dynamical bias governing where contacts occur on the particle surface, arising from shear-induced stress anisotropy and collision dynamics. Here, in contrast, we isolate the purely geometric contribution by asking how any distribution of surface contacts maps into a distribution of contact-normal orientations. Under the assumption of uniform boundary sampling, this mapping reduces to the curvature weighting $P(\theta) \propto 1/\kappa(\theta)$. In this sense, the present model does not attempt to describe the full dynamical process of contact formation, but instead establishes the geometric transform linking boundary geometry to orientational statistics. The close agreement between this minimal construction and both DEM simulations and laboratory measurements suggests that much of the observed nematic alignment may emerge from geometric constraints intrinsic to particle shape, with dynamics acting primarily as a modulation of this underlying geometric bias. This distinction between geometric constraints and dynamical modulation may also help explain why different particle classes exhibit varying degrees of agreement with the smooth-curvature predictions developed here.

This distinction between geometric constraints and dynamical modulation also provides a useful perspective for interpreting the role of particle orientation in the evolution of the contact-normal fabric. Dynamical biases may influence the present framework either through the particle orientation fabric itself or, perhaps more directly, through the contact-normal formulation of fabric considered by Wang *et al.* [28]. In their anisotropic critical state theory, particle orientation enters the evolution of the contact-normal fabric through a negative exponential term that retards the rate at which the contact-normal fabric approaches its steady state. The geometric framework developed here suggests an intriguing possible interpretation of this exponential. Our construction assumes that contacts are sampled uniformly along the particle boundary, such that the resulting distribution of contact-normal orientations is determined entirely by the local boundary curvature. If these contact locations are viewed as a stochastic sequence along the particle boundary, then a uniform sampling process is naturally associated with Poisson statistics, for which the distribution of spatial intervals between successive contacts is exponential. This raises the possibility that the negative exponential appearing in Wang *et al.*'s formulation may reflect, at least in part, the underlying stochastic statistics of contact formation rather than being solely an empirical description of fabric evolution. In this interpretation, the geometric mapping developed here would determine how uniformly sampled contact locations generate the steady-state contact-normal fabric, while Poisson statistics could provide a natural description of the stochastic approach toward that state. Whether such a connection can be derived rigorously, and whether it explains the specific exponential form used in ACST, remains an open question. Of course, the exponential may also have been selected for its analytical advantages: It is positive definite, smoothly modulates the evolution rate, and introduces only a single additional parameter. Distinguishing between these possibilities would provide a compelling direction for future work.

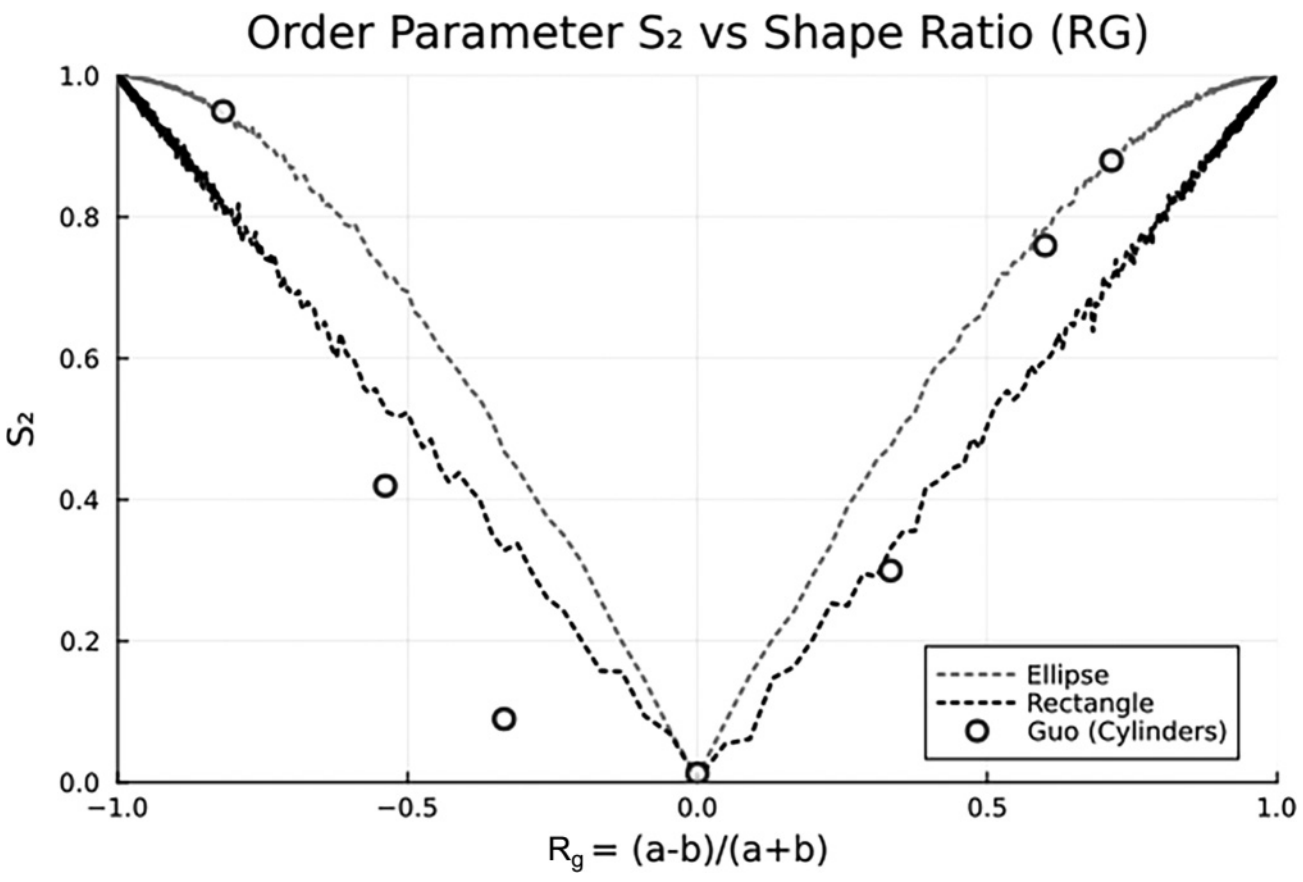

FIG. 4. Geometric model performance across distinct particle classes. Comparison of $S_2$ predictions for smooth ellipsoids (dotted line) and piecewise-flat rectangles (dashed line) against data for cylinders and disks (open circles) from Guo *et al.* [25].

An intriguing observation in our comparison between the geometric models and the DEM data from Guo *et al.* [25] is the transition of cylinders, which possess rectangular cross sections, from following the discrete rectangle prediction at moderate aspect ratios to aligning with the continuous elliptical prediction at larger aspect ratios. We suspect that this alignment with ellipses occurs because of an additional effect where a small tilt out of the shear plane creates an elliptical cross section within the shear plane. For the same angle of tilt, this effect becomes more pronounced for shape values $R_g$ farther from zero. Furthermore, while the potential for planar contacts alters the nature of contact availability and the driving dynamics of the system, in a dense packing of highly elongated rods, flat ends are effectively shielded. This makes planar interactions highly improbable compared to contacts along the elongated sides. Because even a slight out-of-plane tilt effectively introduces nonzero curvature into the shear plane, future work examining the fully three-dimensional kinematics of faceted and planar shapes would be highly valuable. Despite these three-dimensional nuances, the capacity of this framework to capture the behavioral bounds of such extreme geometries points to a much broader applicability of the curvature-based model.

This broader applicability is supported by the findings of Börzsönyi *et al.* [21], who observed robust, shape-dependent macroscopic alignment across diverse laboratory materials—not just in strictly elliptical rice grains, but also in nonelliptical glass cylinders. Furthermore, it is well established that microstructural organization in these systems is fundamentally driven by collision events [22]. Crucially, the orientational bias generated by these collisions is dictated by the distribution of branch vectors and contact normals formed during impact. Because the available contact normals are mapped directly from the geometry of the particle boundary, the fundamental curvature of the particle remains an essential driver of fabric anisotropy.

More broadly, because the present argument is geometric, it should remain relevant across dense particulate flows spanning dry-granular and viscous-suspension regimes, provided the system is sufficiently particle-rich and experiences sustained particle contacts. For instance, in crystal-rich magmas, this same geometric constraint may therefore provide a useful baseline for illuminating the development of crystal fabric and lineation. The data used for validation in this work shows that the quasistatic state is sufficient to meet these conditions. Of future interest would be to explore whether or not it is necessary, as the work of Börzsönyi *et al.* [21] suggest that similar alignment behavior may persist beyond the quasistatic regime. We expect, however, that as inertial effects become increasingly important, geometric controls will progressively lose predictive power.

## VI. CONCLUSION

We have demonstrated that the first-order alignment of nonspherical particles in dense granular flow can be predicted from a purely geometric consideration of particle boundary curvature. By mapping the local curvature to a probability distribution of contact normals, we recover the macroscopic fabric anisotropy observed in complex three-dimensional DEM simulations across a wide range of particle aspect ratios.

Our model assumes uniform contact probability along the perimeter, and its success suggests that geometric constraints are the dominant factor in determining particle orientations contribution to steady-state fabric. This purely geometric baseline successfully reconciles previous theoretical hypothesis of "geometric saturation" in granular contact networks with empirical observations of diverse analog laboratory experiments [21,27]. Future refinements could incorporate kinematic effects, such as the tumbling behavior of particles with extreme aspect ratios, or extend the framework to include nonuniform contact distributions driven by specific flow configurations. Nevertheless, this minimal geometric framework provides a robust baseline for predicting the statistical emergence of granular fabric in multiphase systems.

## ACKNOWLEDGMENTS

Financial support was provided to C.H. by NSF EAR Postdoctoral Fellowship Award No. 2518495. This research was also supported in part by NSF Grant No. PHY-2309135 to the Kavli Institute for Theoretical Physics (KITP), whose unique environment facilitated the origins of this work. E.C.P.B. was supported by the NERC-IRF (NE/V014242/1), the Leverhulme Trust Grant Award No. RPG-2024-294, and the Royal Society (IEC/NSFC/242381). Financial support was also provided to P.Z. by the Leverhulme Trust Grant No. RPG-2024-294.

## DATA AVAILABILITY

The data that support the findings of this study, together with the source code, simulation parameters, and analysis scripts used to generate and analyze those data, are publicly available [30]. The repository includes the workflow required to reproduce Figs. 3 and 4.